\documentstyle[12pt]{article}

\begin{document}
\begin{flushright}
Preprint IHEP 94-56\\
May 1994\\
Submitted to Phys.Lett.B.
\end{flushright}
\vspace*{1cm}
\Large

\begin{center}
{ \bf Double charmed baryon production at $B$-factory.}
\end{center}
\normalsize
\vspace*{0.7cm}

\begin{center}
V.V.Kiselev, A.K.Likhoded and M.V.Shevlyagin
\end{center}

\begin{center}
{\it Institute for High Energy Physics, Protvino, 142284, Russia}
\end{center}

\vspace*{2cm}
{\bf Abstract}

\noindent
The cross-section for the double charmed baryon production at a $B$-factory
is estimated on the basis of the perturbative QCD calculations for the
$c c$-diquark production as well as of the quark-hadron duality.

\newpage
\section{Introduction}
\noindent
Recently a noticeable progress has been achieved in the study of production,
decay and spectroscopy of hadrons, containing two heavy quarks [1-12].
The expected production number of such hadrons with respect to those
with a single heavy quark is of the order of $10^{-(3\div 4)}$. For example,
at the $Z^0$-boson pole the number of events with heavy quarks
is $\sim 10^6$, consequently  the number of hadrons with two heavy quarks
is expected to be $\sim 100-1000$. Taking into account specific decay modes
of hadrons with two heavy quarks one may expect the detection
of single events with such hadrons, which makes their observation at LEP
problematic.

In this letter we consider the double charmed baryon production
($\Sigma^{(*)}_{cc}$) under the conditions of a $B$-factory with
high luminosity $L=10^{34}$~cm${}^{-2}$s${}^{-1}$, where the number of
$\Sigma^{(*)}_{cc}$ is by two orders of magnitude greater than
at the $Z^0$-boson pole.

\section{Fragmentation mechanism}
\noindent
In the recent paper [12] the authors have estimated the production of
$\Sigma_{cc}$-, $\Sigma_{bc}$-, $\Sigma_{bb}$-, $\Lambda_{bc}$-baryons
in the region of the heavy quark fragmentation at high energies.
These estimations are based on the exact analytical calculations
for heavy quarkonium production in
the QCD perturbation theory in the limit of small $m^2/s$ ratio and
nonrelativistic potential model [9,10,11]. In ref.[12]
the diquark $cc$ momentum spectrum has been considered to be equal to
that of heavy vector quarkonium $(c\bar c)$\footnote{In ref. [12]
there is a wrong  additional factor 2.}
\begin{equation}
D_{c\to cc}(z)=\frac{2}{9\pi}\frac{|R_{cc}(0)|^2}{m^3_c}\alpha^2_s(2m_c)F(z)\:,
\end{equation}
where
\begin{equation}
F(z)=\frac{z(1-z)^2}{(2-z)^6}(16-32z+72z^2-32z^3+5z^4)
\end{equation}
and $R_{cc}(0)$ is a radial wave function of the bound diquark at the origin.

Let us note that identical quarks $cc$ in the colour antitriplet state
can only have symmetrical spin wave function in the $S$-wave, i.e.
they must be in the state with the total spin $S=1$.
The normalization of the fragmentation function $D_{c\to cc}(z)$ is determined
by the model dependent value of $R_{cc}(0)$. In ref.[12] a
rather rough approximation with the Coulomb potential in the system of heavy
quarks has been used. This factor gives noticeable uncertainty in
the estimation of the $\Sigma^{(*)}_{cc}$ yield. Moreover, expression (1),
obtained in the scaling limit
$m^2/s \to 0$, is unsuitable for the estimations of the $\Sigma^{(*)}_{cc}$
production at a $B$-factory, where the $m^2/s$ ratio is not small.

Earlier, we have proposed another method to estimate the production of
the hadrons, containing two heavy quarks on the basis of the
quark-hadron duality [9,10].

\section{Calculations under quark-hadron duality}
\noindent
The production cross-section of the $B_c$-meson $S$-wave states at the
$Z^0$-boson pole, calculated in the fragmentation model (1) [9,10,11],
is in a good agreement with the cross-section estimations of the quark pair
$({\bar b}c)$ production in the colour singlet state with small invariant
masses
\begin{equation}
m_b+m_c < M({\bar b}c) < M_{th}=M_B+M_D+\Delta M\:,
\end{equation}
where $\Delta M \simeq 0.5\div 1$~GeV.

In the same range of duality (3) the $(bc)$-``diquark'' production
cross-section is approximately equal to that of $({\bar b}c)$-pair.
Selecting the colour antitriplet state $({ b}c)$ by multiplying
by the factor 2/3, we have  obtained the estimation of
the $\Sigma^{(*)}_{bc}$ and
$\Lambda_{bc}$-baryon production cross-sections on the level
$\sigma (\Sigma^{(*)}_{bc},\;\Lambda_{bc})/\sigma (b{\bar b}) \simeq 6\cdot
10^{-4}$, i.e. 6 times greater than the estimate made in ref. [12] for
the production of $1S$-states. This difference is caused by the fact
that, first,
the contribution of the higher $(bc)$-diquark $nS$-, $nP$-levels must
be taken into the account and, secondly,
the strong suppression is determined by the small value of $R_{bc}(0)$.

Let us consider the $\Sigma^{(*)}_{cc}$-baryon production at the energy of
the $B$-factory ($\sqrt{s}=10.58$~GeV). Note once more that expression (1)
may not be used at the given energy because the power corrections over
$M^2/s$ are substantial. The method of calculations
in the leading order of the QCD perturbation theory  has been
proposed by us in  [9].

In the method of the quark-hadron duality the cross-section for the associated
production of the quarkonium bound states can be estimated using the formula
\begin{equation}
\sum_{nL,J}^{} \sigma (e^+e^- \to \bigl( nL(c{\bar c})_J \bigr)
+c+{\bar c}) =
\int \limits_{m_0}^{M_{th}}
\frac{ d\sigma ( e^+e^- \to (c{\bar c})_{singlet}+c+{\bar c} }
{ dM_{c{\bar c}} } dM_{c{\bar c}}
\end{equation}
where $m_0=2m_c$ --- kinematical threshold of ($c\bar c$)-pair production,
$M_{th}=2M_D+\Delta M$, $\Delta M \simeq 0.5\div 1$~GeV.

\begin{table}[t]
\caption{The $\eta (\psi)$-meson production cross-sections
in $e^+e^-$ annihilation
at the B-factory.}
\begin{center}
\begin{tabular}{|c|c|c|c|c|}    \hline
state  &  $\eta_c(1S)$  &  $\psi (1S)$  &  $\eta_c (2S)$  &  $
\psi (2S)$ \\ \hline
$\sigma$, pb  &  0.025   & 0.055  &  0.003  &  0.010       \\ \hline
\end{tabular}
\end{center}
\end{table}

In Table 1 the results for the numerical calculations of the QCD
perturbation theory diagrams are presented for the production of
the bound $1S$- and $2S$-levels of charmonium at the energy
$\sqrt s$=10.58~GeV and $\alpha_s$=0.2. The values of the radial wave
functions at the origin $R_{nS}(0)$ have been determined from the experimental
data on the lepton decay widths of charmonia $\psi (nS)$ [14]. As it is
evident from Table~1, below the threshold for the decay into
$D\bar D$ mesons the sum over the $S$-wave states of charmonia  is equal to
\begin{equation}
\sigma (\Sigma \eta_c,\psi)=0.093~\mbox{pb}\:.
\end{equation}

Note that the ratio of the vector and pseudoscalar state yields
at the energy $\sqrt{s}=10.58$~GeV is equal to $\omega_V/\omega_P \simeq 2.2$
in contrast to the value $\omega_V/\omega_P \simeq 1$ obtained in the
fragmentation mechanism [12].

Our estimations of the integral in the r.h.s. of expression (4) give
\begin{equation}
\sigma_{c\bar c} (\Delta M=0.5~\mbox{GeV})=0.093~\mbox{pb}\:,
\end{equation}
\begin{equation}
\sigma_{c\bar c} (\Delta M=1~\mbox{GeV})=0.110~\mbox{pb}\:,
\end{equation}
where we set $m_c$=1.4~GeV.

{}From equations (5) and (6) it follows, that the relation of the quark-hadron
duality (4) is well satisfied for the bound states of the ($c\bar c$)-system.

The distributions over the invariant mass of the $c\bar c$-
and $cc$-pair are presented on Figs. 1a and 1b, respectively.
{}From these figures one can see that these
distributions  coincide with each other in the region of small invariant
masses. Hence, in the same duality region, the estimates of
the ($c\bar c$)-pair and  ($cc$)-``diquark'' cross-sections are approximately
equal to each other, so that
\begin{equation}
\sigma_{cc} (\Delta M=0.5~\mbox{GeV})=0.086~\mbox{pb}\:,
\end{equation}
\begin{equation}
\sigma_{cc} (\Delta M=1~\mbox{GeV})=0.115~\mbox{pb}\:.
\end{equation}
Selecting the antitriplet colour state, one can obtain the total
$\Sigma^{(*)}_{cc}$-baryon production cross-section
\begin{equation}
\sigma (\Sigma^{(*)}_{cc}) \simeq (70\pm 10)\cdot 10^{-3}~\mbox{pb},
\end{equation}
thus the relative number of double charmed baryons is equal to
\begin{equation}
\sigma (\Sigma^{(*)}_{cc})/\sigma (c{\bar c}) \simeq 7\cdot 10^{-5}.
\end{equation}

For the luminosity $L=10^{34}\mbox{cm}^{-2}\mbox{s}^{-1}$ the number
of events with the $\Sigma^{(*)}_{cc}$ production  is equal to
$N(\Sigma^{(*)}_{cc})\simeq 7\cdot 10^3$ per year, so it is by two orders
of magnitude greater than the yield of the $\Sigma^{(*)}_{cc}$ baryons at LEP.

The distribution over the ($cc$)-diquark momentum at the
asymmetric collider at KEK [13] (8$\times$3.5~GeV) is presented on Fig. 2.

\section{Conclusion}
\noindent
In the present letter we have made  exact calculations of the double
charmed
$\Sigma^{(*)}_{cc}$-baryon production in the leading order of
QCD perturbation theory and on the basis of the quark-hadron duality.
We present the $\Sigma^{(*)}_{cc}$ production cross-section at the energy
$\sqrt{s}=10.58$~GeV of the $B$-factory, where the fragmentation formula [12]
is not applicable.

The main theoretical uncertainty in the estimations for the production
cross-section of the double charmed baryons is related
with the description of the
process for the heavy $(cc)$-diquark hadronization. First of all, a
considerable fraction of the diquarks (1/3)
is produced in the colour sixtet state
and can fragment into both the exotic four quark states ($cc{\bar q}{\bar q}$)
and the  $DD$-meson pair. As in ref. [12] we assume that colour
antitriplet
state hadronizes into the $\Sigma^{(*)}_{cc}$-baryon with  100\%
probability. Thus, at the $B$-factory one could expect $10^4$ events per year
of the  $\Sigma^{(*)}_{cc}$-baryon production
at the luminosity $L=10^{34}~\mbox{cm}^{-2}\mbox{s}^{-1}$.

\vspace*{1.cm}
In conclusion, the authors thank S.S.Gershtein for interesting discussions.
The work of M.V.Shevlyagin was supported by the Russian
Fund of Fundamental Researches (project number 93-02-14456).

%\newpage

%%%%%%%%%%%%%%%%%%%%%%%%%%%%%%%%%%%%%%%%%%%%%%%%%%%%%%%%%%%%%%%%%%%%
\newpage
\section*{Figure Captions}
\begin{description}
\item[Fig.1] The distributions over: a) the invariant mass $M_{c\bar c}$ of
$c\bar c$-pair,\\  b) the invariant mass $M_{cc}$ of $cc$-pair.
\item[Fig.2] The distribution over the ($cc$)-diquark momentum.
\end{description}

\end{document}